\documentclass[sigconf]{acmart}

\AtBeginDocument{%
  }

\setcopyright{none}
\renewcommand\footnotetextcopyrightpermission[1]{}
\acmConference[ESSC'24]{ESWEEK 2024 Embedded System Software Competition}{September 29 -- October 4 2024, Raleigh, NC, USA}

\usepackage{cleveref}

\usepackage[acronym]{glossaries} 
\newacronym{soa}{SoA}{State of the Art}

\begin{document}

\title{ATOMIC: Automatic Tool for Memristive IMPLY-based Circuit-level Simulation and Validation}


\author{Fabian Seiler}
\orcid{0009-0000-6517-451X}
\affiliation{
\institution{TU Wien, Institute of Computer Technology}
\city{Vienna}
\country{Austria}
}
\email{fabian.seiler@student.tuwien.ac.at}

\author{Nima TaheriNejad}
\orcid{0000-0002-1295-0332}
\affiliation{
\institution{Institute of Computer Engineering, Heidelberg University}
\city{Heidelberg}
\country{Germany}
}
\email{nima.taherinejad@ziti.uniheidelberg.de}

\renewcommand{\shortauthors}{Trovato et al.}

\begin{abstract}
Since performance improvements of computers are stagnating, new technologies and computer paradigms are hot research topics. Memristor-based In-Memory Computing is one of the promising candidates for the post-CMOS era, which comes in many flavors. Processing In memory Array (PIA) or using memory, is on of them which is a relatively new approach, and substantially different than traditional CMOS-based logic design. Consequently, there is a lack of publicly available CAD tools for memristive PIA design and evaluation.
Here, we present \textbf{ATOMIC}: an \textbf{A}utomatic \textbf{To}ol for \textbf{M}emristive \textbf{I}MPLY-based \textbf{C}ircuit-level Simulation and Validation.
Using our tool, a large portion of the simulation, evaluation, and validation process can be performed automatically, drastically reducing the development time for memristive PIA systems, in particular those using IMPLY logic. The code is available at
\textbf{https://github.com/fabianseiler/ATOMIC}.
\footnote{
This manuscript was written as a technical document corresponding to the ATOMIC project. This project was submitted and presented at the Embedded System Software Competition (ESSC) 2024 at ESWEEK. This project was developed for two recent papers~\cite{seiler2024SSAxIMC, seiler2024NoCarry}. It was made public to present the community with a useful tool for memristive PIA design. If you have found this project useful or included it in another project please refer to this manuscript and the papers it was developed for.}
\end{abstract}


\maketitle

\section{Motivation}

\begin{figure}[tb]
    \centering
    \includegraphics[width=1\columnwidth]{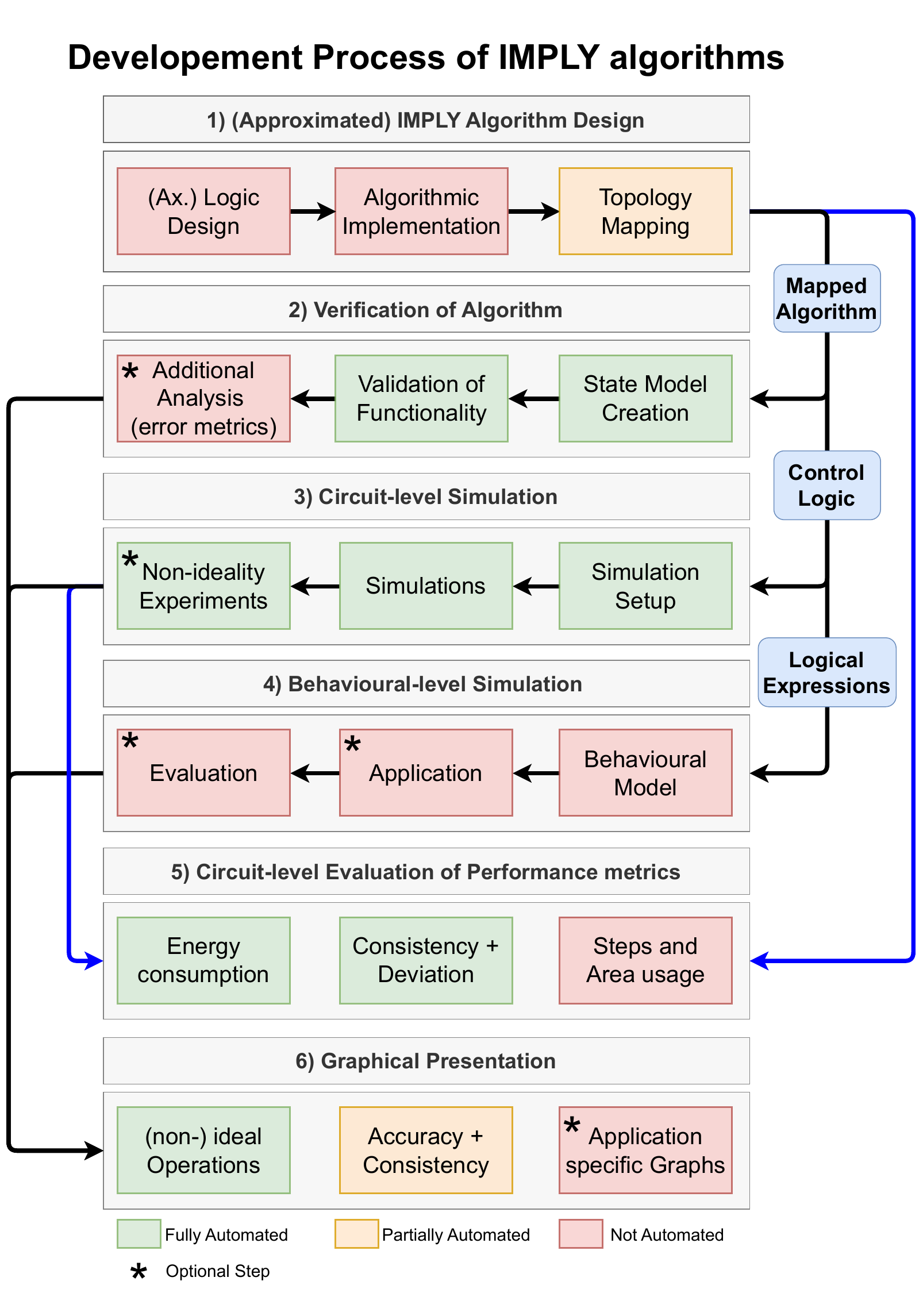}
    \caption{Design process for (exact and approximated) algorithms in memristive IMPLY logic. The blocks in green are fully, and the blocks in orange are partially automated in this project.}
    \label{fig:Dev_process}
\end{figure}

With the stagnating computer performance, there is a rising emphasis on new computing technology and paradigms such as In-memory Computing (IMC) with memristors and the approximation of calculations or Approximate Computing (AxC). 
Memristor-based In-Memory Computing is one of the promising candidates for the post-CMOS era, which comes in many flavors. Processing In memory Array (PIA) or using memory, is on of them which is a relatively new approach, and substantially different than traditional CMOS-based logic design. Consequently, there is a lack of publicly available CAD tools for memristive PIA design and evaluation. 
We illustrated the typical development process for (approximated) algorithms for IMPLY, a common logic form for memristive computing, in \Cref{fig:Dev_process}. 
When carried out manually, the process can take multiple hours to days, even for experienced researchers. \cite{seiler2024NoCarry}

Since there is a lack of available tools that help automate the development we propose \textbf{ATOMIC}, an automated Python tool based on the PyLTSpice framework. With this tool, a large portion of the validation, simulation, and evaluation process, as well as the processing and illustration of the data, is now completely automated. 
Since real-world memristors experience non-idealities such as variation, studying their effect is an important topic. The deviation of the memristor's resistive state is one of the most important variations that is often disregarded in many \gls{soa} papers. Therefore, we placed a strong emphasis on this topic to facilitate simulation and evaluation of memristive circuits under such variations. 
We approached this project in a generic fashion to allow compatibility with exact and approximated algorithms in various topologies and provide an environment that can be easily expanded to beyond what is presented here. 

\begin{figure*}[t]
    \centering
    \includegraphics[width=0.9\textwidth]{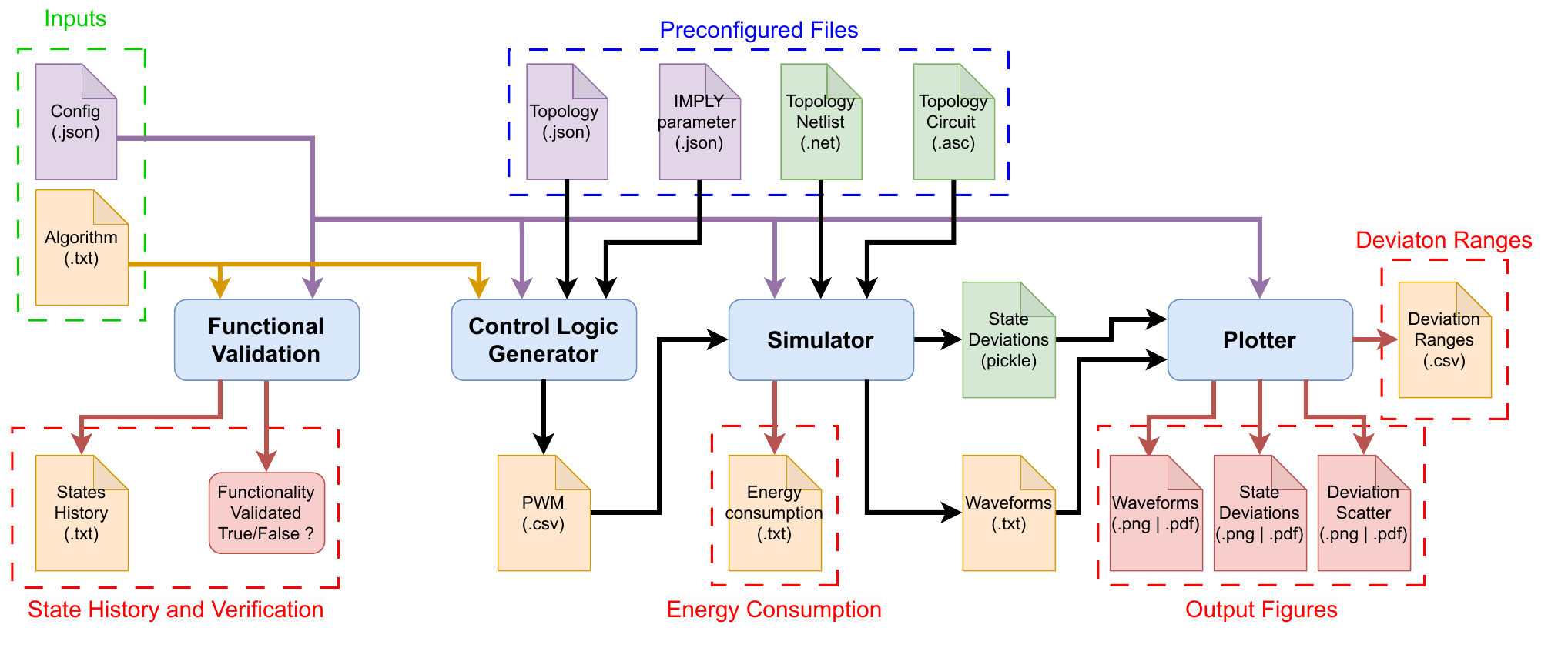}
    \caption{Overview of the ATOMIC pipeline and dataflow.}
    \label{fig:Project_Overview}
\end{figure*}

\section{Overview of ATOMIC}

In this section, we will give an overview of the project structure and briefly explain the functions of the individual classes. The project was written in Python and
consists of four classes: Functional Validation, Control Logic Generator, Simulator, and Plotter. We also implemented a logger in a singleton design pattern that is embedded in all classes to store information vital for debugging. An overview of the information and process flow is illustrated in \Cref{fig:Project_Overview}.

\subsection{Files}
\subsubsection{Inputs Files}
For this project, two files have to be configured beforehand. The first one is the ``config'' file where the essential information of the algorithm, used topology, required memristors, and expected outputs are stored in JSON format. The exact sequence of the algorithm has to be stored in an additional plain-text file. The specific format for each of the implemented topologies will be explained in more detail in \Cref{Setup}.

\subsubsection{Pre-configured Files}
We stored important information such as the memristors, switches, and corresponding control voltages for each implemented topology in JSON files. Since the simulations are based on SPICE, we prepared circuit (.asc) and netlist (.net) files for each topology.
The IMPLY-specific parameters are configured with values commonly used in the \gls{soa}. They can be adjusted quickly in the ``IMPLY\_parameters.json'' file. All of the pre-configured files can be found in the ``Structures'' folder.

\subsubsection{Outputs Files}
Since ATOMIC automatically evaluates many steps in the design process we store intermediate and final results to allow for easy debugging. All intermediate and final results are stored in sub-folders of the ``outputs'' folder. More details on the individual result files can be found in the class description that creates them.

\subsection{Classes}
\label{classes}

\subsubsection{Functional Validation}
The first class in the pipeline is responsible for the creation of a state model and the validation of the functionality for the given algorithm. From the configuration file, the number of inputs and outputs are extracted, and logic vectors that represent each input combination are initialized. With the method \texttt{calc\_algorithm}, each line of the algorithm files is read and processed. Depending on the topology, this method extracts information from each line and creates either IMPLY or FALSE operations. These operations are then applied to the state model in a vectorized fashion via \texttt{imply\_op} and \texttt{false\_op}. The operation and updated state model are stored in the ``State\_History.txt'' file for later examination. At the end of the algorithm, the equivalence of the expected and simulated logic vectors is checked.

\subsubsection{Control Logic Generator}
This class is responsible for generating control logic in the form of PWM signals, that are then applied in the SPICE simulation. When initialized the class creates ``.csv'' for each memristor and switch that is used in the given algorithm and stores them in the ``PWM\_output'' folder. The method \texttt{eval\_algo} iterates over every step in the algorithm file and stores time steps for each memristor and switch individually. The parameters like cycle time and control voltages for different operations are read from the ``IMPLY\_parameters.json'' file. After the last step the complete PWM signal is written in the previously created files. 

\subsubsection{Simulator}
The simulator class is responsible for manipulating netlists, simulations with LT-SPICE, and calculating the energy consumption. When initialized, the parameters of the configuration and the topology file are extracted, and the netlist is selected with the \texttt{SpiceEditor} class from the PyLTSpice framework. With the \texttt{set\_parameters} method, a list of parameters in the format 
\texttt{param\_values:= list(memristor values, R\_on, R\_off) } is accepted as an input and the netlist is manipulated accordingly. But instead of changing our preset netlist, we store the intermediate netlist and the other manipulated simulation files in a temporary folder to allow for better debugging. With the method \texttt{run\_simulation} a transient (as configured) simulation with SPICE is started. The resulting waveforms can be extracted with \texttt{read\_raw} and saved as a file with \texttt{save\_raw}. The energy consumption is calculated with \texttt{calculate\_energy}. With the method \texttt{evaluate\_deviation} we automated resistive deviation experiments that calculate different input combinations. For each input, the state deviation of memristors is varied, and the results are stored in the ``Waveforms'' folder. Since we are interested in the state of the memristors at the end of the algorithm, we store the resulting logic states of the output memristors in the ``deviation\_results'' folder. This is also done with different combinations of resistive deviation.

\subsubsection{Plotter}
\begin{figure*}[t]
    \centering
    \includegraphics[width=0.85\textwidth]{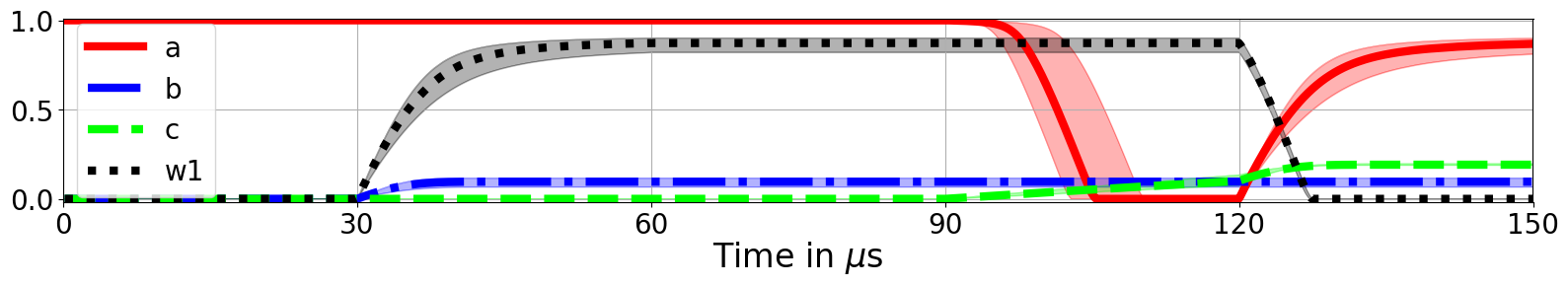}
    \caption{Example waveform with a deviation of $\pm 20\%$ illustrated as the shaded area.}
    \label{fig:Waveform}
\end{figure*}

\begin{figure*}[t]
    \centering
    \includegraphics[width=0.9\textwidth]{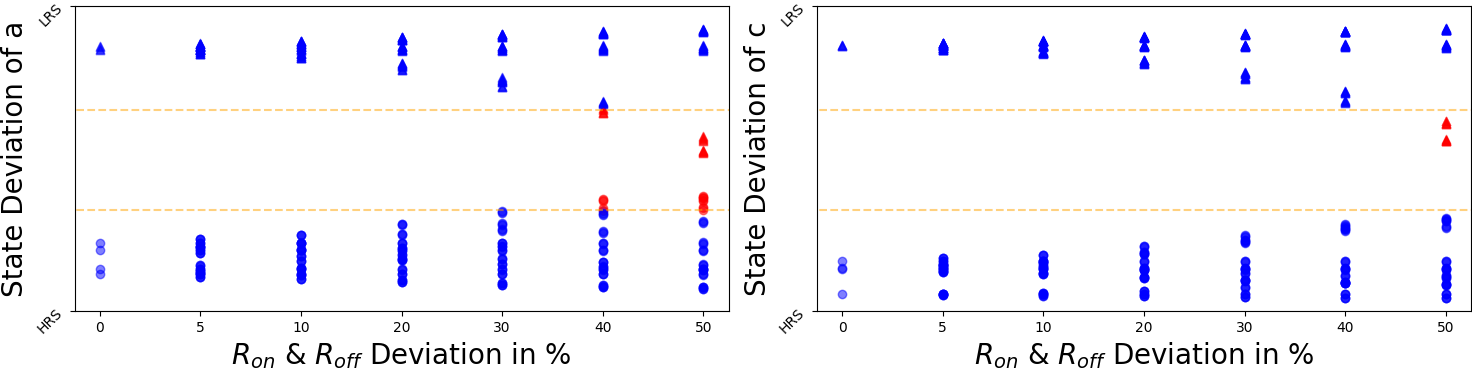}
    \caption{Scatter plot of output states with increasing deviation range. The markers in red are marked as incorrect results.}
    \label{fig:Scatter}
\end{figure*}

\begin{figure}[tb]
    \centering
    \includegraphics[width=0.55\columnwidth]{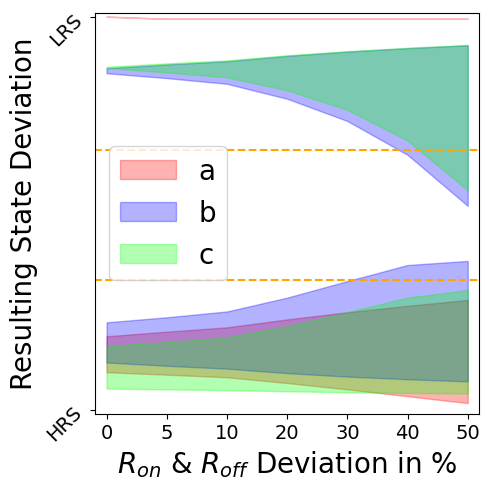}
    \caption{Range of the resulting states for each output over increasing deviation.}
    \label{fig:Range}
\end{figure}

This class is responsible for post-processing the simulation data and illustrating the results. The state of individual memristors in an IMPLY algorithm can vary depending on the resistive state and its deviation. To visualize the impact of these deviations on the algorithm we implemented the method \texttt{plot\_waveforms\_with\_deviation} that extracts the range of the waveforms and plots the result. An example is shown in \Cref{fig:Waveform}, where the deviation range is illustrated as the shaded area around the exact waveform.
To analyze the functionality of algorithms with increasing deviation, the method \texttt{plot\_deviation\_scatter} was implemented. An example plot is shown in \Cref{fig:Scatter}, where the incorrect output states are colored red. \texttt{plot\_deviation\_range} calculates the range of the different outputs and compares them over increasing deviation. An example is shown in \Cref{fig:Range}. The resulting figures are stored in the folder ``Images''. To compare with other algorithms the output state ranges are stored in the file ``deviation\_range.txt''.

\section{How to use ATOMIC}
\label{Setup}

\subsection{Requirements}
This project was developed for Python 3.12, LT-SPICE version 17.1.6.0, and PyLTSpice version 5.3.1.
The required Python packages are summarized in the ``requirement.txt'' to allow for a fast setup.

\subsection{Setup algorithm file}
Since there is no uniform way of writing down IMPLY-based algorithms, we propose a simple and easily extendable format for storing algorithms in various topologies. Our framework is currently compatible with the serial, semi-serial, and semi-parallel topologies but can be extended quite easily.  
Each section (depending on the topology) that can compute operations must be set to either FALSE, IMPLY, or NOP (No Operation) and the sections have to be separated by the ``|''  symbol. 
The FALSE operation can reset up to three different memristors, which is written in the form \textbf{F$j$} \textbf{F${j,k}$} or \textbf{F${j,k,l}$} where $j$, $k$, and $l$ correspond to the number each used memristor is given. The IMPLY operation is written as \textbf{I${j,k}$}, which is equal to the operation $M_{k}' = M_j \rightarrow M_k$ where memristor $j$ implies memristor $k$.
The memristor number resembles an enumeration of a list of used memristors (e.g. ["a", "b", "c", "w1"] have the numbers [0, 1, 2, 3] so the operation ``\textbf{I}0,2'' is equal to $c' = a \rightarrow c$). We implemented a few \gls{soa} IMPLY algorithms to give the user a reference on how the algorithms can be written in each topology.

\subsection{Setup config file}
The configuration file has to be created separately for each algorithm. It contains information on the algorithm's intended behavior and declares the use case of each memristor. Additionally, the expected output states have to be declared to allow for validation of the algorithm's functionality via the state model or on circuit-level. We kept this configuration as general as possible to also allow approximated algorithms in various forms. We prepared a template here and in the project files to speed up the preparation. The unfilled template can be seen here:

\begin{verbatim}
{
    "topology": "",    
    "algorithm": "",
    "memristors": ["", "", "", ""],
    "inputs": ["", "", ""],                          
    "work": [""],                             
    "outputs": ["", ""],                        
    "switches": ["", "", "", ""],
    "steps": ,
    "output_states": {"": [0,0,0,0,0,0,0,0],
                      "": [1,1,1,1,1,1,1,1]}
}
\end{verbatim}

\noindent The topology name must be either be ``Serial'', ``Semi-Serial'', or ``Semi-Parallel'' and the algorithm should be the name of the prepared algorithm file. In ``memristors'' the name of all used memristors must be declared. If they are used as input, work, or output they have to be declared in the corresponding place. We note here that memristors can be used for both input and output, as well as work and output, as this is a common design property of \gls{soa} IMPLY algorithms. In ``switches'' each used switch has to be declared, as they can vary between algorithms. More information on the available switches and memristors for each topology can be found in the ``Structures'' folder. ``output\_states'' is a dictionary that includes the name of the outputs and the expected bit vectors. The number of outputs is variable to allow for a more flexible design, which is necessary in this design space. More information and example algorithms can be found in the project.

\subsection{Run the pipeline}
We implemented two variants on how an algorithm can be evaluated. The first one is a Jupyter Notebook (\texttt{Pipeline.ipynb}) where each step of the pipeline can be executed individually. We suggest this method for the initial stages of the development so possible mistakes can be better detected. 
When only parts of the project are required we refer the reader to \Cref{classes} for more information on the implemented classes.
To run the whole pipeline (algorithm validation, simulation, deviation experiments, and illustration) we prepared the command:

\texttt{python Pipeline.py -- config\_file=CONFIG\_FILENAME.json}

\noindent 
We configured additional flags so the user can customize various settings, which can be seen in the project.
We implemented a few \gls{soa} algorithms and corresponding configuration files, to give some examples. To evaluate all the implemented algorithms at once, run the command: 
\texttt{python evalute\_soa.py}

\section{How to extend this project}
\subsection{Appending topologies or other structures}
We built this project in a highly modular fashion to allow for easy extensibility. 
As interest in this area of research is increasing we also expect many new topologies and algorithms to be published. We marked areas where new topologies can be added in all implemented classes. It is also required to create a new circuit with the used naming convention for all circuit elements and to parameterize them like the implemented topologies.
To append new topologies or other memristive structures to the functional validation module, add a new branch in \texttt{FunctionalValidation.calc\_algorithm()} that converts the algorithm to IMPLY and FALSE commands.
Since the control logic can differ drastically between topologies, the creation of PWM signals is highly specific. We recommend that a new method:
\texttt{ControlLogicGenerator.write\_timestep\_TOPOLOGY}\\ 
is created and the specifics are handled there. We again marked the spaces where branches should be added in all methods.
As the simulator only manipulates the netlist and runs SPICE commands, not much change is required for new topologies. We note here that the \texttt{calculate\_energy()} and \texttt{evaluate\_deviation} functions are limited to three inputs by design since we only evaluated adder circuits. To run simulations with arbitrary inputs first prepare a list of parameter values and then utilize the \texttt{run\_simulation} and \texttt{save\_raw} methods to simulate certain inputs and store the resulting waveforms.
In the plotter module all methods are generalized for an arbitrary number of output states. We note here that the figure size may have to be adjusted. 

\subsection{Other logic forms and memristor model}
There exist various forms of logic for memristive circuits. To add other logic forms, the operational functions of the Functional Validation class must be extended to subject to the new logic. As the Control Logic Generator class is designed for IMPLY, another class that handles the generation of PWM signals would be the best option. The Simulator and Plotter can handle arbitrary logic forms.

To utilize another model for memristors, the sub-module in the selected topology has to be exchanged. As these models use different parameters, the method \texttt{Simulator.set\_parameters()} and all the dependent methods must also be updated accordingly. \cite{seiler2024NoCarry}

\glsresetall

\bibliographystyle{ACM-Reference-Format}
\bibliography{references}

\end{document}